# Sub-cycle attosecond control in frustrated double ionization of molecules with orthogonally polarized two-color laser fields

A. Vilà,[1] G. P. Katsoulis,[1] and A. Emmanouilidou[1]

[1]*Department of Physics and Astronomy, University College London, Gower Street, London WC1E 6BT, United Kingdom*

(Dated: May 21, 2018)

We demonstrate sub-cycle control of frustrated double ionization (FDI) in the two-electron triatomic molecule $D_3^+$ when driven by two orthogonally polarized two-color laser fields. We employ a three-dimensional semi-classical model that fully accounts for the electron and nuclear motion in strong fields. We control FDI triggered by a strong near-infrared laser field with a weak mid-infrared laser field. This control as a function of the time delay between the two pulses is demonstrated when the FDI probability and the distribution of the momentum of the escaping electron along the mid-infrared laser field are considered in conjunction. We find that the momentum distribution of the escaping electron has a hive-shape with features that can accurately be mapped to the time one of the two electrons tunnel-ionizes at the start of the break-up process. This mapping distinguishes consecutive tunnel-ionization times within a cycle of the mid-infrared laser field.

## I. INTRODUCTION

Frustrated double ionization (FDI) is a major process in the nonlinear response of multi-center molecules when driven by intense laser fields, accounting roughly for 10% of all ionization events [1, 2]. In frustrated ionization an electron first tunnel-ionizes in the driving laser field. Then, due to the electric field of the laser pulse, it is recaptured by the parent ion in a Rydberg state [3]. This process is a candidate for the inversion of $N_2$ in free-space air lasing [4]. In FDI an electron escapes and another one occupies a Rydberg state at the end of the laser pulse. FDI has attracted considerable interest in recent years in a number of experimental studies in the context of $H_2$ [1] and of the triatomic molecules $D_3^+$ and $H_3^+$ [5–7].

Two pathways were identified to underlie FDI in previous theoretical studies of strongly-driven two-electron diatomic and triatomic molecules [2, 8]. Electron-electron correlation is important, primarily, for one of the two pathways. Recently, we proposed a road for future experiments to identify the important role of electron-electron correlation in FDI [9]. We employed a triggering 800 nm strong laser field and an orthogonally polarized probing 400 nm weaker laser field to control the relevant pathway for electron-electron correlation in FDI. We showed that, together, the FDI probability and the V-shape of the momentum distribution of the escaping electron along the fundamental laser field bear clear signatures of the turning on and off of electron-electron correlation in FDI.

Here, we demonstrate sub-cycle attosecond control in FDI of triatomic molecules by employing orthogonally polarized two-color laser fields (OTC). In particular, we employ a near-infrared (near-IR) 800 nm laser field and a week mid-IR 2400 nm laser field. We show that the FDI probability changes significantly as a function of the time delay between the two laser fields. The change is mainly due to the FDI pathway where electron-electron correlation is important. Interestingly, we find that the momentum distribution of the escaping electron along the mid-IR laser field has a striking hive-shape. We show that there is a one-to-one correspondence between the features of this hive-shape and the time that one of the two electrons tunnel-ionizes at the start of the break-up process of the strongly-driven molecule. For the laser field parameters we consider here, these tunnel-ionization times take place around the extrema of the near-IR laser field. The above mentioned mapping allows us to distinguish FDI ionization from consecutive tunnel-ionization times within a cycle of the mid-IR laser field.

Two-color laser fields are an efficient tool for controlling electron motion [10, 11] and for steering the outcome of chemical reactions [12–14]. Other applications include the field-free orientation of molecules [15–17], the generation of high-harmonic spectra [18–21] and probing atomic and molecular orbital symmetry [22–24]. It was recently demonstrated experimentally that a weak mid-IR laser pulse acts as a streak camera that time-resolves the strong-field dynamics of the escaping electron in single ionization triggered by a near-IR pulse [25] in atoms.

Our work employs a three-dimensional (3D) semi-classical model. Classical and semi-classical models are essential for understanding the break-up dynamics of strongly-driven triatomic molecules [8, 26]. One reason is that treating two electrons and three-nuclei in a strong laser field is a challenge for fully ab-initio quantum mechanical calculations. The latter techniques can currently address one electron triatomic molecules [27]. Our 3D model has provided significant insights into FDI for strongly-driven $H_2$ [2] and $D_3^+$ [8]. Moreover, our previous result for the distribution of the kinetic energy release of the Coulomb exploding nuclei in FDI of $D_3^+$ agreed well with experiment [7].

## II. METHOD

The OTC laser field we employ consists of an 800 nm laser field, i.e. $\omega_1 = 0.057$ a.u., with field strength of $E_{\omega_1}$ equal to 0.08 a.u. and a weak 2400 nm laser field, i.e. $\omega_2 = \omega_1/3$ with field strength $E_{\omega_2}$ equal to 0.0253 a.u. These field strengths correspond to the intensity of the mid-IR field being one tenth of the intensity of the near-IR field. The ponderomotive energies, $E^2/(4\omega^2)$, of the near-IR and mid-IR laser fields are equal to 0.49 a.u. and 0.44 a.u., respectively, and are hence comparable. In our previous studies with OTC laser fields we employed a probing field of 400 nm with small ponderomotive energy of 0.048 a.u. [9] The combined OTC laser field is

$$\mathbf{E}(t, \Delta t) = E_{\omega_1} f(t)\cos(\omega_1 t)\hat{z} + E_{\omega_2} f(t - \Delta t)\cos(\omega_2(t - \Delta t))\hat{x}$$
$$f(t) = \exp\left(-2\ln 2 \left(\frac{t}{\tau}\right)^2\right), \quad (1)$$

with $\tau = 40$ fs the full-width-half-maximum duration.

For our current studies, we employ the initial state of $D_3^+$ that is accessed experimentally via the reaction $D_2 + D_2^+ \to D_3^+ + D$ [5, 7]. It consists of a superposition of equilateral triangular-configuration vibrational states $\nu = 1 - 12$ [7, 28]. We assume that most of the $D_3^+$ ionization occurs at the outer classical turning point of the vibrational levels [29, 30]. The turning point varies from 2.04 a.u. (v = 1) to 2.92 a.u. (v = 12) [28, 31]. We initialize the nuclei at rest for all vibrational levels, since an initial pre-dissociation does not significantly modify the ionization dynamics [32].

The combined strength of the two laser fields is within the below-the-barrier ionization regime. Hence, we assume that one electron (electron 1) tunnel-ionizes at time $t_0$ in the field-lowered Coulomb barrier. For this quantum-mechanical step, we compute the ionization rate using a semi-classical formula [33]. $t_0$ is selected using importance sampling [34] in the time interval the two-color laser field is present. The ionization rate is then used as the importance sampling distribution. For electron 1, the velocity component that is parallel to the OTC laser field is taken equal to zero while the transverse one is given by a Gaussian [35]. The initial state of the initially bound electron (electron 2) is described by a microcanonical distribution [36].

Another quantum mechanical aspect of our 3D model is tunneling of each electron during the propagation with a probability given by the Wentzel-Kramers-Brillouin approximation [2, 32]. This aspect is essential to accurately describe the enhanced ionization process [37, 38]. In EI, at a critical distance of the nuclei, a double potential well is formed such that it is easier for an electron bound to the higher potential well to tunnel to the lower potential well and subsequently ionize. The time propagation is classical, starting from time $t_0$. We solve the classical equations of motion for the Hamiltonian of the strongly-driven five-body system, while fully accounting for the Coulomb singularities [32].

To record the FDI events of $D_3^+$, we propagate the trajectories to the asymptotic time limit. The final fragments in FDI are a neutral excited fragment $D^*$, two $D^+$ ions and one escaping electron. In the neutral excited fragment $D^*$ the electron occupies a Rydberg state with quantum number n > 1. We identify two pathways of frustrated double ionization. Their main difference is how fast the ionizing electron escapes following the turn on of the laser field [2]. In pathway A, electron 1 tunnel-ionizes and escapes early on. Electron 2 gains energy from an EI-like process and tunnel-ionizes. It does not have enough drift energy to escape when the laser field is turned off and finally it occupies a Rydberg state, $D^*$. In pathway B, electron 1 tunnel-ionizes and quivers in the laser field returning to the core. Electron 2 gains energy from both an EI-like process and the returning electron 1 and tunnel-ionizes after a few periods of the laser field. When the laser field is turned off, electron 1 does not have enough energy to escape and remains bound in a Rydberg state. It follows that electron-electron correlation is more pronounced in pathway B [2].

We compute the FDI probability as a function of the time delay $\Delta t$ of the $\omega - \omega/3$ laser pulses using

$$P^{FDI}(\Delta t) = \frac{\sum_{\nu, i} P_\nu \Gamma(\Delta t, \nu, i) P^{FDI}(\Delta t, \nu, i)}{\sum_{\nu, i} P_\nu \Gamma(\Delta t, \nu, i)}, \quad (2)$$

where i refers to the different orientations of the molecule with respect to the z-component of the laser field. We consider only two cases of planar alignment, that is, one side of the equilateral, molecular triangle is either parallel or perpendicular to the $\hat{z}$–axis. $\Gamma(\Delta t, \nu, i)$ is given by

$$\Gamma(\Delta t, \nu, i) = \int_{t_i}^{t_f} \Gamma(t_0, \Delta t, \nu, i) dt_0, \quad (3)$$

where the integration is over the duration of the OTC field. $\Gamma(t_0, \Delta t, \nu, i)$ is the ionization rate at time $t_0$ for a certain molecular orientation i, vibrational state $\nu$ and time delay $\Delta t$. $P_\nu$ is the percentage of the vibrational state $\nu$ in the initial state of $D_3^+$ [28]. $P^{FDI}(\Delta t, \nu, i)$ is the number of FDI events out of all initiated classical trajectories for a certain molecular orientation i, vibrational state $\nu$ and time delay $\Delta t$. Due to the challenging computations involved, we approximate Eq. (2) using the $\nu = 8$ state of $D_3^+$, see also [9]. This is a reasonable approximation, since we find that the $\nu = 6, 7, 8$ states contribute the most in the sum in Eq. (2). These states yield very similar results for the FDI probabilities and the distributions of the momentum of the escaping electron.

## III. RESULTS AND DISCUSSION

In Fig. 1, we plot the FDI probability and the probability of pathway A and B as a function of the time delay between the mid-IR and the near-IR laser pulses.

The time delay is expressed in units of the period of the near-IR laser field, since the FDI probability is periodic with a period of $T_{\omega_1}/2$. We consider $\Delta t$ in the time interval $[-1.5, 1.5)T_{\omega_1}$. One cycle of the mid-IR laser field corresponds to three cycles of the near-IR laser field. We find that the FDI probability changes as a function of the time delay. This change is mainly due to the significant change of the probability of pathway B which varies from 2.6% to 0%. In contrast, the probability of pathway A only varies from 1.1% to 0.77%. In the absence of the mid-IR laser field the probability of pathway A and B is 3.6% and 4.9%, respectively. Thus, the probing mid-IR laser field decreases the maximum value of the FDI probability from 8.5% to 3.7%. In our previous study, the probing 400 nm laser pulse did not affect the maximum value of the FDI probability [9]. The difference between the two studies is that the ponderomotive energy due to the triggering 2400 nm laser field is large while the one due to the triggering 400 nm laser field was small.

Another consequence of the large ponderomotive energy of the mid-IR laser pulse is that the FDI probabilities at a consecutive maximum and minimum decrease with increasing values of $\Delta t$ in the time interval $[-1.5, 1.5)T_{\omega_1}$, see Fig. 1. We find that the overall values of the FDI probabilities are larger for negative time delays. Electron 1 tunnel-ionizes mostly from the extrema of the near-IR laser field around the extremum corresponding to its peak intensity. For negative time delays, the atom encounters first the peak of the mid-IR and then the peak of the near-IR laser pulse. Therefore, for negative (positive) time delays when electron 1 tunnel-ionizes it mostly encounters a decreasing (increasing) force from the mid-IR laser field along the $\hat{x}$-axis. This force results in both electrons moving away from the nuclei and thus in an increase of the double ionization probability (not shown) and a decrease of the FDI probability. Since this force decreases with increasing $\Delta t$ it follows that the FDI probability decreases with increasing time delay.

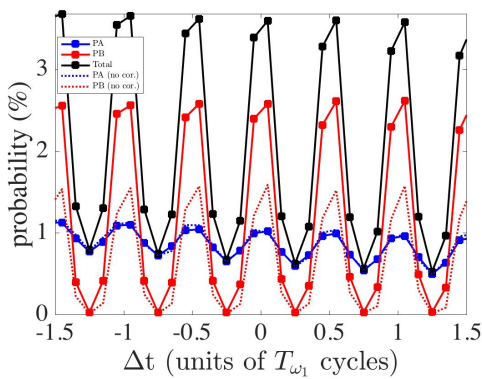

FIG. 1. FDI probability and probabilities of pathways A and B as a function of the time-delay with electron-electron correlation turned on (solid lines) and off (dotted lines).

Electron-electron correlation plays a significant role for pathway B of FDI [2, 9]. In this work we show this to be the case by turning on and off the electron-electron correlation in our computations, see Fig. 1. We find that the probability of pathway A as a function of the time delay is not affected by the electron-electron correlation being turned on or off. However, the maximum value of the probability of pathway B when electron-electron correlation is turned off reduces to only half the value it has when electron-electron correlation is on. Hence, electron-electron correlation is important for pathway B.

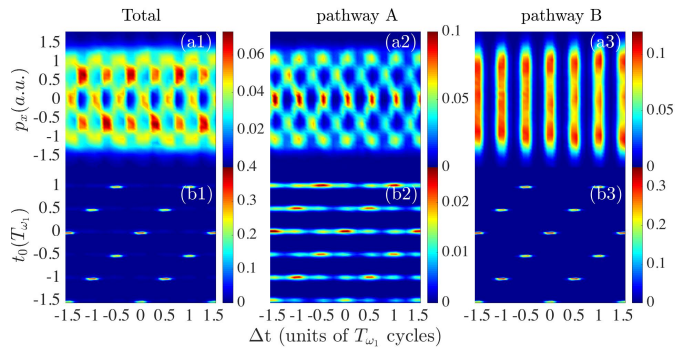

FIG. 2. The distribution of $p_x$ for FDI (a1) and for pathways A (a2) and B (a3) are plotted as a function of $\Delta t$. For each $\Delta t$, the distribution of $p_x$ in (a2)-(a3) is normalized with respect to the total FDI probability, while in (a1) is normalized to 1. The distribution of the time electron 1 tunnel-ionizes for FDI (b1) and for pathways A (b2) and B (b3) is plotted as a function of $\Delta t$. For each $\Delta t$, the distribution of $t_0$ in (b1)-(b3) is normalized with respect to the total FDI probability.

In Fig. 2(a1), we plot the distribution of $p_x$ of the escaping electron for FDI as a function of $\Delta t$ in the time interval $[-1.5, 1.5)T_{\omega_1}$ in steps of $\Delta t = 0.1T_{\omega_1}$. We find that the distribution of $p_x$ has a hive-shape. This hive-shape is mainly due to pathway A, see Fig. 2(a2). To understand the shape of $p_x$ for pathway A, we consider separately the contribution of the mid-IR laser field, $\Delta p_x^E$, and of the core $\Delta p_x^C$ to the final momentum $p_x$ with $p_x = \Delta p_x^E + \Delta p_x^C + p_{x,t_i}$. $\Delta p_x^C$ is the momentum change due to the nuclei as well as the electron-electron interaction, but we find the latter contribution to be very small. $p_{x,t_i}$ is the x-component of the momentum of the escaping electron at time $t_i$. For pathway A $t_i$ is the time that electron 1 tunnel-ionizes, $t_0$. We find that $p_{x,t_0}$ has only a small contribution to $p_x$. The momentum change from the mid-IR laser field and the core are given by

$$\Delta p_x^E(\Delta t, t_i) = \int_{t_i}^{\infty} -E_{\omega_2}(t) dt, \qquad (4)$$

$$\Delta p_x^C(\Delta t, t_i) = \int_{t_i}^{\infty} \left( \sum_{i=1}^{3} \frac{\mathbf{R_i} - \mathbf{r_1}}{|\mathbf{r_1} - \mathbf{R_i}|^3} + \frac{\mathbf{r_1} - \mathbf{r_2}}{|\mathbf{r_1} - \mathbf{r_2}|^3} \right) \cdot \hat{x} dt.$$

In Fig. 3(c) we plot $\Delta p_x^C$ as a function of the time delay. $\Delta p_x^C$ has a two-band structure that is symmetric with respect to $p_x$ equal to zero. We find that the upper (lower) band corresponds mostly to FDI events where electron 1 tunnel-ionizes from the negative (positive) x̂-axis.

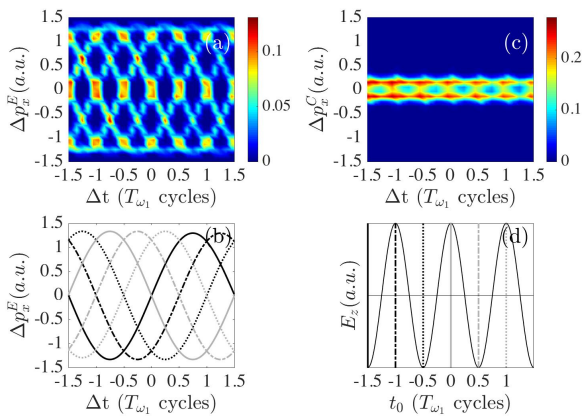

FIG. 3. The distributions of $\Delta p_x^E$ (a) and $\Delta p_x^C$ (c) for pathway A of FDI are plotted as a function of $\Delta t$. For each $\Delta t$ the distributions in (a) and (c) are normalized to the probability of pathway A of FDI. (b) $\Delta p_x^E$ is plotted as a function of $\Delta t$ for the six $t_0$s corresponding to extrema of $E_{\omega_1}$ within one cycle of the 2400 nm laser field, which are shown in (d).

Fig. 3(a), clearly shows that the hive-shape of the distribution $p_x$ is accounted for by $\Delta p_x^E$ of electron 1. To elucidate $p_x$ for pathway A we first investigate the time electron 1 tunnel-ionizes as a function of $\Delta t$, see Fig. 2(b2). When the mid-IR laser field is turned off $t_0$ is centered around the extrema of the near-IR laser field. For one cycle of the mid-IR laser field electron 1 tunnel-ionizes from six extrema of the near-IR laser field, see Fig. 3(d). We find that when the mid-IR laser field is turned on electron 1 still tunnel-ionizes from these six extrema. However, tunnel-ionization is favored a bit more from extrema where the combined OTC laser field is larger and thus the ionization rate is higher. Indeed, we compute for each $\Delta t$ the time $t_{max}$ when the combined OTC field in Eq. (1) is maximum. We find that it coincides with $t_0$ for pathway A in Fig. 2(b2). This is expected since when electron 1 is the escaping electron the time electron 1 tunnel-ionizes will be roughly equal with the time the ionization rate is maximum.

Given the above, in Fig. 3(c), we reproduce the outline of the hive-shape of $p_x$ in Fig. 3(a)—the result of our full scale computations—by employing a very simple model. Specifically, using Eq. (1), we compute $\Delta p_x^E$ as a function of $\Delta t$ for each of the six $t_0$s electron 1 tunnel-ionizes from (Fig. 3(d)). Fig. 3(b) clearly shows how the cos/sin like curves of $\Delta p_x^E$ as a function of $\Delta t$ for each of the six $t_0$s within one cycle of the mid-IR laser field intertwine to result in the hive-shape of $p_x$ for pathway A. Hence, we have established a one-to-one correspondence, i.e. mapping, between the features of the hive-shape in Fig. 3(a) and in Fig. 2(a2) and the times electron 1 tunnel-ionizes within one cycle of the mid-IR laser pulse. That is, the 2400 nm laser pulse, probes and separates the momentum $p_x$ that corresponds to different $t_0$s. This mapping distinguishes the FDI probability from different $t_0$s within one cycle of the mid-IR laser field but not between $t_0$s that differ by an integer number of cycles.

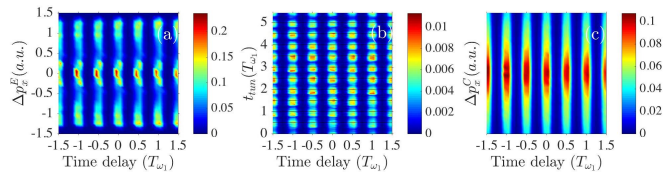

FIG. 4. The distributions of $\Delta p_x^E$ (a), the time that electron 2 tunnel-ionizes during the time propagation, $t_{tun}$, (b) and $\Delta p_x^C$ for pathway B of FDI (c) are plotted as a function of $\Delta t$. For each $\Delta t$ the distributions in (a)-(c) are normalized to the probability for pathway B of FDI.

We next focus on the contribution of pathway B to the distribution of $p_x$, see Fig. 2(a3). First, we explain why the probability of pathway B is zero at $\Delta t = (n+1/4)T_{\omega_1}/2$. We do so using the six $t_0$s at the extrema of the near-IR laser field in the interval $[-1.5, 1.5)T_{\omega_1}$ where electron 1 tunnel-ionizes. Fig. 3(b) shows that at $\Delta t = (n+1/4)T_{\omega_1}/2$ $\Delta p_x^E$ is not zero for any of the six $t_0$s. When the momentum change of electron 1 due to the mid-IR laser field is not zero then electron 1 does not return to the core. As a result, electron 1 can not transfer energy and ionize electron 2 and the probability for pathway B goes to zero. However, $\Delta p_x^E$ of electron 1 is equal to zero for two out of the six $t_0$s at $\Delta t = nT_{\omega_1}/2$, see Fig. 3(b), resulting in non zero probability of pathway B. These two $t_0$s correspond to the two extrema of the mid-IR laser field within one cycle, where the vector potential of the mid-IR laser field is zero.

In Fig. 4(a) and (c) we plot the momentum change of the escaping electron for pathway B due to the mid-IR laser field and due to the nuclei plus the electron-electron repulsion, respectively. To compute these contributions, we employ Eq. (4) using as $t_i$ the time $t_{tun}$. The latter is the time electron 2 tunnel-ionizes during the time propagation. Unlike pathway A, the contribution of the core for pathway B is broad at the $\Delta t$s where the probability of pathway B is not zero, see Fig. 4(c). This is consistent with electron 2 being the escaping electron in pathway B, since electron 2 has more time to interact with the nucleus before it tunnel-ionizes and finally escapes.

Moreover, Fig. 4(a) shows that the momentum change of electron 2 from the mid-IR laser field is similar to $p_x$ of electron 1 for pathway A at the $\Delta t$s where the probability of pathway B is not zero. Indeed, for pathway A we showed that $\Delta p_x^E$ of electron 1 for the six $t_0$s in the time interval $t_0 \in [-1.5, 1.5)T_{\omega_1}$ gives rise to the hive-shape of the distribution $p_x$. This is also the case for pathway B, however, the relevant $t_i$ time in Eq. (4) is not $t_0$ but the time $t_{tun}$ electron 2 tunnel-ionizes during the time propagation. We show in Fig. 4(b) that electron 2 tunnel-ionizes around the extrema of the near-IR laser field. Therefore, the distribution of $p_x$ of electron 2 for pathway B has a hive-shape as does the distribution of $p_x$ of electron 1 for pathway A. The main difference is that pathway B is zero at $\Delta t = (n+1/4)T_{\omega_1}/2$. Another difference is that $t_{tun}$ is more broadly distributed than $t_0$ around the extrema of the near-IR laser field. As a result, the hive-shape for pathway B is more broad than pathway A at $\Delta t = nT_{\omega_1}/2$. Thus, $p_x$ for pathway B contributes to the hive-shape of the total FDI distribution $p_x$.

Here, we focused on sub-cycle control. However, we note that our current results also demonstrate control of electron-electron correlation in FDI, as did our previous results in ref. [9]. In the current study, the hive-shape of the momentum distribution of the escaping electron along the mid-IR laser field is due to both pathways A and B at $\Delta t = nT_{\omega_1}/2$. At $\Delta t = (n+1/4)T_{\omega_1}/2$, the probability of pathway B is zero and the FDI probability is reduced. The electron momentum distribution and the FDI probability as a function of the time delay are experimentally accessible observables. Thus, when taken in conjunction, the change of each of these two observables with time delay is a way of asserting experimentally the presence of electron-electron correlation in FDI.

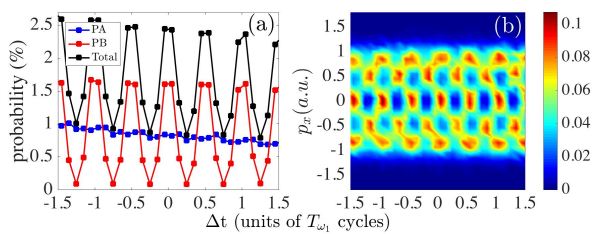

FIG. 5. (a) and (b) the same as Fig. 1 and Fig. 2(a1), respectively, for $H_2$.

Finally, we show that sub-cycle attosecond control can also be achieved with OTC fields for $H_2$. We choose $E_{\omega_1}=0.064$ a.u. so that $E_{\omega_1}$ for $H_2$ and $D_3^+$ has the same percentage difference from the field strength that corresponds to over-the-barrier ionization. We choose $E_{\omega_2}$ to be such that the intensity of the $E_{\omega_2}$ laser field is one tenth of the intensity corresponding to the $E_{\omega_1}$ laser field, as for the $D_3^+$ molecule. We show in Fig. 5(a) that the FDI probability changes from a maximum value of 2.7% to a minimum of 1% as a function of the time delay between the two pulses. The probability of pathway A remains almost constant with a change from 1% to 0.9%. The change of probability of pathway B from 1.7% to 0.1% is mainly responsible for the change in the FDI probability. However, we note that in Fig. 5 we only consider the orientation where the inter-nuclear axis is parallel to the 800 nm laser field since for the perpendicular orientation the probability is zero. Therefore, the FDI probability averaged over all molecular orientations will be smaller than the values presented in Fig. 5(a). Smaller FDI probability aside, Fig. 5(b) shows that for $H_2$ the hive-shape of the momentum distribution of the escaping electron for FDI as a function of $\Delta t$ is similar to the one for $D_3^+$.

## IV. CONCLUSIONS

In conclusion, we have shown that sub-cycle attosecond control for the FDI process can be achieved with OTC laser fields in $D_3^+$. We employ a near-IR laser field to trigger the FDI process and a weak mid-IR laser field to probe and control FDI. We find that the FDI probability changes sharply with the time delay between the two laser fields. Moreover, we identify a hive-shape in the momentum distribution of the escaping electron along the mid-IR laser field. This hive-shape is mapped back to tunnel ionization of one of the two electrons, at the start of the molecular break-up process, from six consecutive extrema of the near-IR laser field within one cycle of the mid-IR field. Moreover, we have shown that pathway B is non zero only when tunnel-ionization takes place from extrema of the near-IR laser field that coincide with extrema of the mid-IR laser field. These extrema correspond to zero momentum change of the escaping electron from the mid-IR laser field. Future studies could explore the interference of electrons escaping with the same momentum but tunnel-ionizing from different extrema of the near-IR laser field.

*Acknowledgments.* A.E. acknowledges the EPSRC grant no. J0171831 and the use of the computational resources of Legion at UCL.